\documentclass[12pt,a4paper]{article}
\usepackage{amsmath,graphics}
\begin{document}

\begin{center}
{\huge Spin Injection: Interface Resistance in Fe/Semiconductor
  Junctions Calculated from First Principles}

\vspace{1cm}

Olaf Wunnicke\footnote{Institut f\"ur Festk\"orperforschung, 
    Forschungszentrum J\"ulich, D-52425 J\"ulich, Germany}, Phivos
    Mavropoulos$^1$,  and Peter H. Dederichs$^1$
\end{center}

\begin{abstract}
  We calculate the current spin polarisation and the interface
  resistance of Fe/GaAs and Fe/ZnSe (001) spin injection junctions
  from first principles, including also the possibility of a Schottky
  barrier. From our results of interface resistance we estimate the
  barrier thickness needed for efficient spin injection if the process
  is non-ballistic.
\end{abstract}

\vspace{1cm}


\noindent
{\bf Running title:} \\
Spin Injection in Fe/Semiconductor Junctions \\

\noindent
{\bf Corresponding author:} \\
Olaf Wunnicke, Institut f\"ur Festk\"orperforschung,
Forschungszentrum J\"ulich, D-52425~J\"ulich, Germany \\ 
email: o.wunnicke@fz-juelich.de \\
tel.: +49-2461-61-4799, fax: +49-2461-61-2620 \\

\newpage

The achievement of spin-polarised electric transport via the conduction
band of semiconductors (SC) is one of the central issues in the field
of spintronics. Potential technological applications have already been
proposed since 1990 \cite{Datta90}, but the conditions under which
this can be realised are not yet clear, neither experimentally nor
theoretically.  The bottleneck seems to be the injection of the
spin-polarised current from a ferromagnetic metal (FM) contact into
the SC.  One basic reason for this was revealed by Schmidt and
collaborators \cite{Schmidt00}, who argued that the huge conductivity
mismatch of the FM and the SC, in correlation with the much smaller
spin-flip mean free path in the FM, will lead to a drastic drop of the
spin polarisation of the current in the vicinity of the interface.
There are, in theory, two ways to circumvent the obstacle: the
ballistic spin injection and the injection through a spin-selective
tunneling barrier. At least the latter has already been used in praxis
with considerable success \cite{Zhu01}.

The possibility of ballistic spin injection was proposed by Grundler
\cite{Grundler01}. First-principles calculations
\cite{Wunnicke02,Mavropoulos02,Zwierzycki02} have shown the necessity
to consider the full band structure of the FM and SC together with the
electronic structure of the interface. Then the difference in symmetry
of the FM and SC for the minority-spin wave functions can play the most
important role, by essentially disallowing the incoming FM minority
electron to pass onto a SC state. Spin polarisations and
magnetoresistance ratios of up to the ideal 100\% were calculated for
Fe/GaAs(001), Fe/ZnSe(001), and Fe/InAs(001) systems.

In parallel to this, there is also the proposal of Rashba
\cite{Rashba00} and Fert and Jaffr\`es \cite{Fert01} that a tunneling
barrier at the FM/SC interface could solve the problem even in the
diffusive regime. The idea is that a tunneling barrier can be highly
spin selective, as shown in TMR junctions, and at the same time
provides an extremely high interface resistance; this combination of
properties would lead to spin injection even in the troubling
diffusive regime.

In this contribution we extend our previous work \cite{Wunnicke02} to
the evaluation of interface resistance for majority and minority
electrons in Fe/GaAs and Fe/ZnSe (001) junctions, with or without a
tunneling barrier, in the approach developed in Ref.~\cite{Schep97},
and compare to the limits set by the Fert-Jaffr\`es theory.  Finally,
we discuss the importance of interface resonances in such junctions.

Schep and collaborators \cite{Schep97}, working in the approximation
of the resistor model, have deduced the resistance of a ballistic
interface of area $S$ between two bulk regions A (in our case Fe) and
B (SC) in which conduction channels are randomly mixed due to diffuse
scattering, as
\begin{equation}
SR = S\frac{h}{e^2}
\left[
\frac{1}{T_{\mathrm{tot}}} - \frac{1}{2}
\left(\frac{1}{N_{\mathrm{Fe}}}+\frac{1}{N_{\mathrm{SC}}}\right)
\right].
\label{eq1}
\end{equation}
Here, $T_{\mathrm{tot}}$ is the ballistic transmission probability of
the interface. The diffusion is assumed to occur only in the bulk
regions. $N_{\mathrm{Fe}}$ and $N_{\mathrm{SC}}$ represent the number
of Landauer conduction channels in the two bulk regions, {\it
  i.e.}~they are proportional to the area of the projected Fermi
surface on the interface plane.

We calculate the electronic structure of the bulk and interface
regions in the local density approximation of density-functional
theory, using the KKR Green function method, and assuming two
independent spin channels. From the Green function the ballistic
scattering properties of the interface are obtained using a method
equivalent to the Landauer-B\"uttiker formalism \cite{Baranger89}. The
scattering at the interface is assumed specular, {\it i.e.}~the part
of the Bloch $\mathbf{k}$-vector which is parallel to the interface,
$\mathbf{k}_{\parallel}$, is conserved; the presupposition for this is
that the interface is defect-free. The transmission probability can be
viewed then as a function of $\mathbf{k}_{\parallel}$, and
$T_{\mathrm{tot}}$ is an integral of $T(\mathbf{k}_{\parallel})$ over
the surface Brillouin zone SBZ:
\begin{equation}
T_{\mathrm{tot}} = \frac{S}{(2\pi)^2}
\int_{\mathrm{SBZ}} d\mathbf{k}_{\parallel}\, 
T(\mathbf{k}_{\parallel})
\label{eq2}
\end{equation}

In order to achieve spin injection, the SC conduction band edge $E_c$
must be slightly lower than the Fermi level $E_F$; in our calculations
we model such a situation by rigidly lowering the SC potential in the
region far from the interface. We use a value of $E_F =
E_c+10\mathrm{meV}$ as a typical one for an experimental situation.
The potential close to the interface, up to two monolayers (ML), is
determined by the self-consistent electronic structure and the
metal-induced gap states. From this point (3rd ML) on the transition
to the bulk-like region would depend on the doping and the exact band
offset of the SC, with the possibility of a longer or shorter Schottky
barrier. We model the situation by either a gradual lowering of the
potential up to the final bulk value, thus describing a Schottky
tunneling barrier, or by an abrupt lowering of the potential to the
final bulk value; more details are given in Ref.~\cite{Wunnicke02}.
The former situation is relevant for the case of spin injection
through a tunneling barrier with an extremely high interface
resistance and spin selectivity. As explained in
Refs.~\cite{Wunnicke02,Mavropoulos02}, for the evaluation of the
current spin polarisation the calculation of
$T(\mathbf{k}_{\parallel})$ at $\mathbf{k}_{\parallel}=0$ is enough
for tunneling or thermal spin injection; for the interface resistance
the integration of eq.~(\ref{eq2}) must be employed, but we note that it
can be limited within a small circle around
$\mathbf{k}_{\parallel}=0$, because of the tiny Fermi sphere in the
SC.

In the semiconductor part, the low value of $E_F-E_c$ means that the
corresponding Fermi sphere is extremely small; then the conducting
channels of the SC are much fewer than the ones in Fe, and in
eq.~(\ref{eq1}) $1/N_{\mathrm{SC}}+1/N_{\mathrm{Fe}}\simeq
1/N_{\mathrm{SC}}$.  Next, one should compare $1/N_{\mathrm{SC}}$ with
$1/T_{\mathrm{tot}}$. In the case of a tunneling barrier,
$T_{\mathrm{tot}}$ is expected to be so small that $1/N_{\mathrm{SC}}$
will be negligible and the interface resistance will be dominated by
the ballistic specular scattering.  This remains true even in the
absence of a tunneling barrier, because $T(\mathbf{k}_{\parallel})$ is
of the order of $0.1$ for majority spin electrons (and orders of
magnitude less for minority) \cite{Wunnicke02}, making
$T_{\mathrm{tot}}$ of the order of $0.1\times N_{\mathrm{SC}}$ or
less.

The results obtained for Fe/ZnSe and Fe/GaAs are presented in Fig.~1.
In the upper plots the current spin polarisation is shown, as a
function of barrier thickness, for all possible terminations of the SC
part. The high degree of polarisation is due to the symmetry mismatch
of the incoming minority Fe states, as explained in
Ref.~\cite{Wunnicke02}.  The values obtained for the polarisation by
use of eq.~(\ref{eq2}) are somewhat smaller than at
$\mathbf{k}_{\parallel}=0$, because at nonzero
$\mathbf{k}_{\parallel}$ the symmetry mismatch for minority is not
strict.

In all cases except the Se terminated Fe/ZnSe system, the current spin
polarisation decreases for thicker Schottky barriers; this shows
mostly in the Zn terminated Fe/ZnSe system. The cause can be traced
back to the existence of resonant interface states in the vicinity of
the Fermi level for the minority spin. As discussed in
Ref.~\cite{Wunnicke02b}, they can contribute to an increase of the
tunneling current. The density of states for the Fe/ZnSe interface at
$\mathbf{k}_{\parallel}=0$ is shown in Fig.~2. For the Zn termination,
the interface state lies clearly in the vicinity of the Fermi level,
while for the Se termination it lies higher and does not contribute to
the conductance.  However, we must note that effects that are not
taken into account here, such as lattice relaxations, a different
pinning of the band offset, or a finite bias, could move the resonant
states away from or onto the Fermi level. In the latter case the
polarisation is strongly affected and can even changes sign, if the
peak is at $E_F$.

In the lower plots of Fig.~1, the interface resistance is presented
for various barrier thicknesses. Note the logarithmic scale used for
the resistance. For ZnSe the constant slope reflects the exponential
decay of the wave functions within the barrier. For GaAs the asymptotic
behaviour of an exponential decay presents itself for thicker
barriers, because of the smaller band gap of this material.

In order to estimate the range where spin injection with a barrier is
most efficient, we take the criterion proposed in Ref.~\cite{Fert01}.
This reads for the case of spin injection and detection via a second
SC/FM interface,
\begin{equation}
r_N\frac{t_N}{l^{\mathrm{sf}}_N}<SR<r_N\frac{l^{\mathrm{sf}}_N}{t_N}
\end{equation}
with $r_N$ the SC resistance (normalised to interface area), $t_N$ its
thickness (between the barriers) and $l^{\mathrm{sf}}_N$ the spin-flip mean
free path. Typically, $r_N$ can be of the order of $10^{-9}\Omega
\mathrm{m}^2$ and $l^{\mathrm{sf}}_N$ of the order of a $\mu\mathrm{m}$
\cite{Fert01}. According to the results presented in Fig.~1, this
places the necessary Schottky barrier thickness in the range of 70\AA\ 
for ZnSe or 100\AA\ for GaAs, if $t_N$ is of the order of a
$\mu\mathrm{m}$.

In conclusion, we have performed ab initio calculations of the
transmission probability in Fe/ZnSe and Fe/GaAs spin injection devices
taking into account the possibility of a Schottky barrier. We have
estimated the corresponding interface resistances for several barrier
thicknesses and found relevant barrier thicknesses for efficient spin
injection, and shown that resonant interface states can lower the spin
injection efficiency. Our presentation is relevant for situations
where diffusive scattering is present in the bulk.

\newpage

\section*{Captions}

\noindent
Fig. 1: Current spin polarization (upper plots) and interface
resistance 
(lower plots) of Fe/ZnSe(001) (left panel) and Fe/GaAs(001)
(right panel), as functions of the Schottky barrier thickness.
The solid curves refer to the cases of Zn and Ga termination,
and the dashed to Se and As termination at the interface. The
values are obtained by an integration over the whole SBZ. Note
the different scales in the $y$ axes. \\

\noindent
Fig. 2: Local DOS of the Fe/ZnSe(001) at $\mathbf{k}_{\parallel}=0$ 
with a Zn (upper plot) and a Se (lower plot) terminated
interface.  The filled gray lines denote the DOS of bulk Fe, the
solid lines the DOS for the interface Fe layer, and the dashed
lines the DOS of the first ML of ZnSe (Zn ML for the upper plot
and Se ML for the lower).  The vertical lines indicate the gap
region in the ZnSe and in between is the position of the Fermi
level. In this plot, the potential is in the ground state
position with no Schottky barrier inserted.

\newpage

\begin{figure}
  \begin{center}
    \resizebox{11cm}{!}{\includegraphics{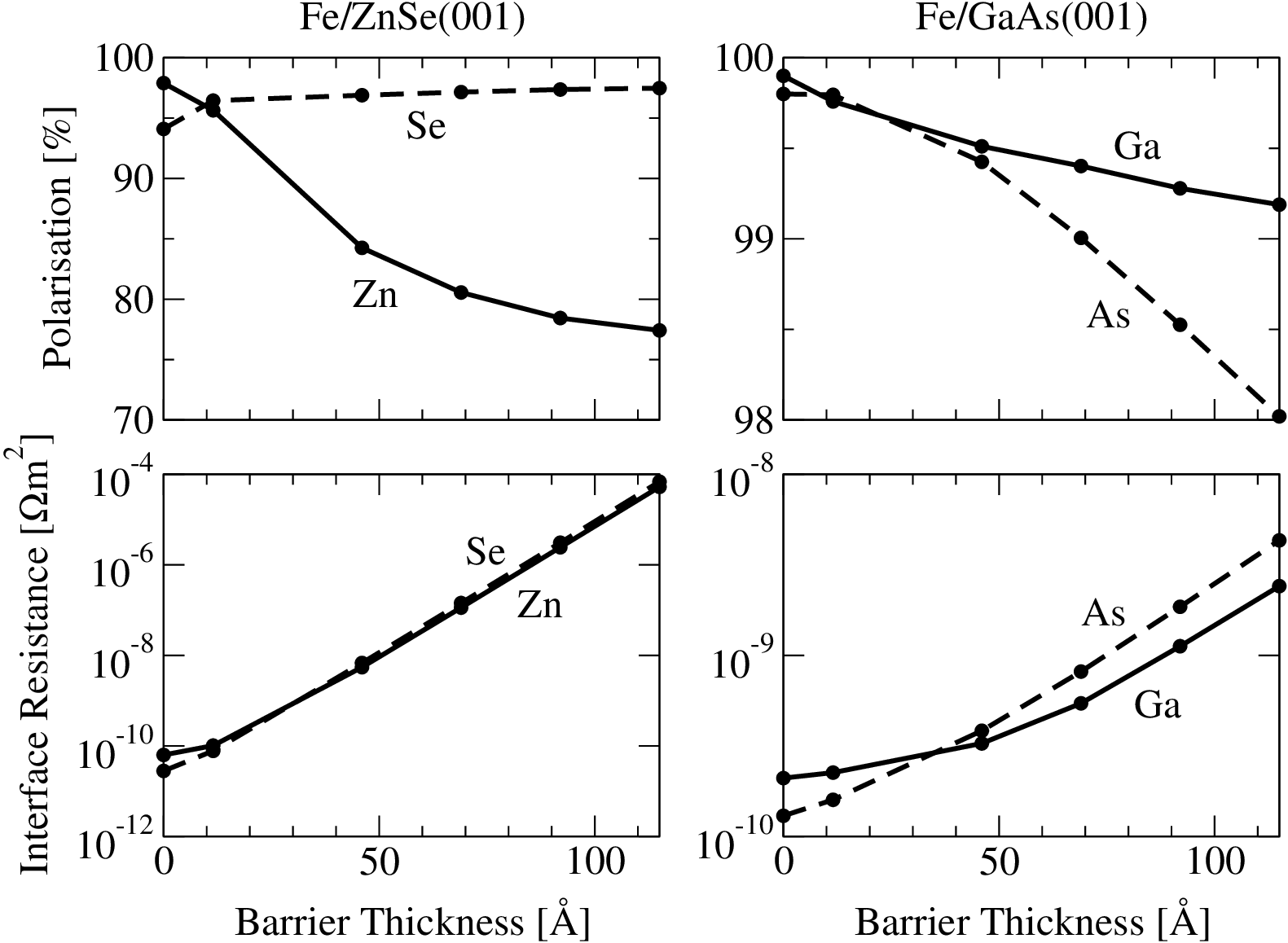}}
    \caption{}
  \end{center}
\end{figure}
\begin{figure}
  \begin{center}
    \resizebox{11cm}{!}{\includegraphics{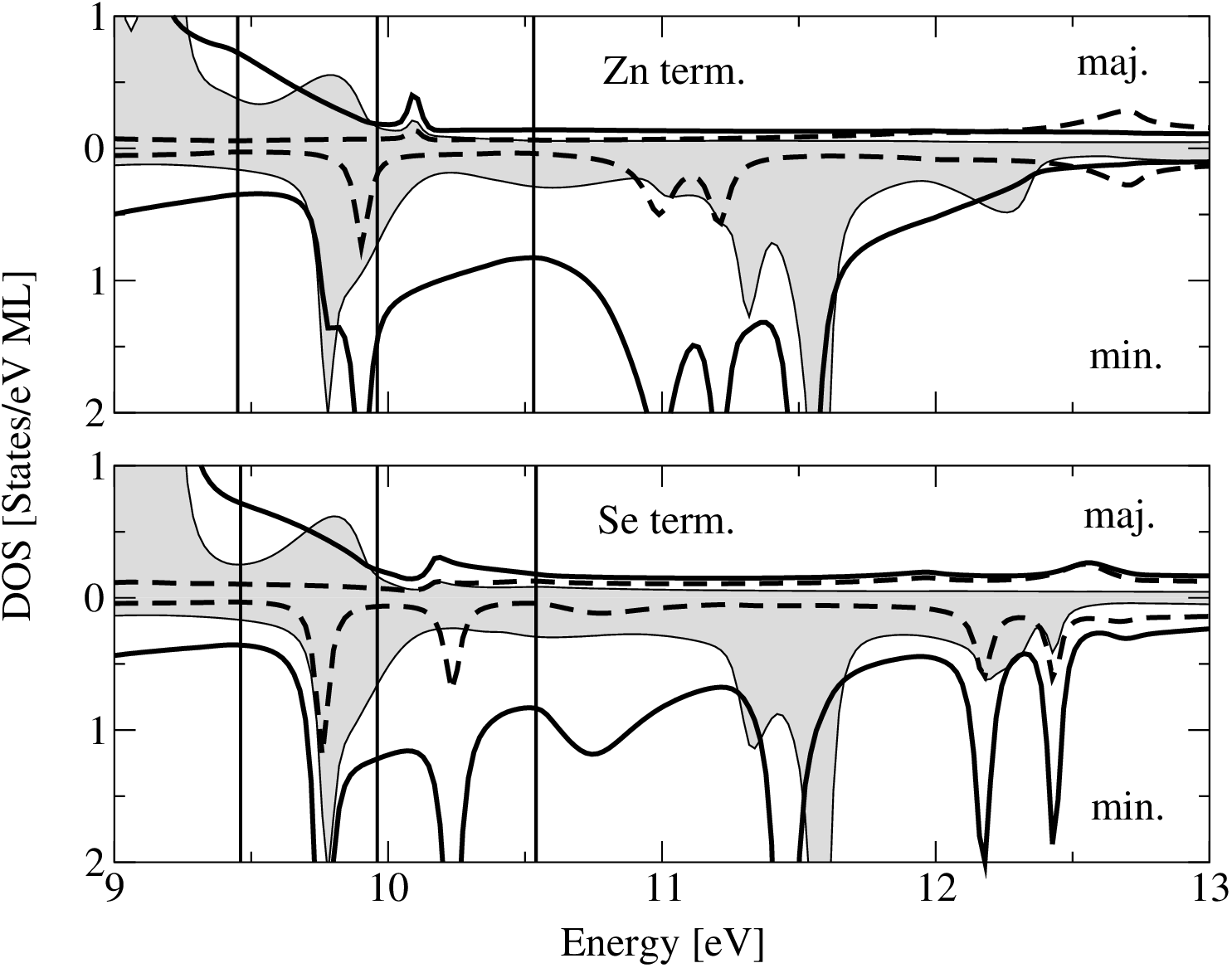}}
    \caption{}
  \end{center}
\end{figure}

\end{document}